\documentclass[twocolumn,showpacs,preprintnumbers,amsmath,amssymb]{revtex4}

\usepackage{graphicx}
\usepackage{dcolumn}
\usepackage{bm}

\begin{document}

\preprint{Zeno}

\title{Revisiting Zeno's paradox: Flying arrows for atom-diatom reactions}

\author{C. Sanz-Sanz}%
\author{A. S. Sanz}%
\author{T. Gonz\'alez-Lezana}\email{tglezana@iff.csic.es}%
\author{O. Roncero}%
\author{S. Miret-Art\'es}%
\affiliation{Instituto de F\'{\i}sica Fundamental-CSIC,
Serrano 123, 28006 Madrid, Spain%
}%

\date{\today}

\begin{abstract}

The possibility to observe quantum Zeno and anti-Zeno scenarios for atom-diatom reactive collisions is investigated for two diferent processes (F+HD and H+O$_2$) by means of 
time-dependent wave packet propagations. A novel approach is proposed in which the survival probabilities investigated are those
obtained when the initial state is redefined after time intervals $\tau$ at which measurements are
performed on the system. The comparison with the actual probability for the unperturbed system
reveals the existence of a regime in which the decay 
from the initial state appears to be inhibited (the quantum Zeno effect) 
or accelerated (the anti-Zeno effect). 
In contrast with preceding interpretations, given that the time-evolving wave packet is not affected
at any time (``{\it flying arrow}"), the present procedure does not require to invoke to 
counterintuitive hindered evolutions (``{\it stopping arrow}") to explain the results.
On the contrary the consecutive update of the reference state of the system solves the apparent modern
version of Zeno's paradox.   
\end{abstract}

\pacs{Valid PACS appear here}
\maketitle


In the extreme limit of one of his paradoxes, Zeno proposed many centuries ago the abolition of motion. The Greek philosopher defied the common sense when he concluded that a flying arrow was at rest at each point of time during its motion to the target.
In order to solve the apparent controversy, one is then forced to invoke the differential calculus by defining the velocity of the arrow, not by its instantaneous position in space, but via the corresponding time derivative of the spatial coordinates. 
The spirit of this postulate has been adopted nowadays in the investigation of
the evolution of a quantum unstable
system when frequent measurements are performed on it. 
According to the quantum Zeno effect (QZE) and anti-Zeno effect (AZE), one can either inhibit or accelerate,
respectively, the decay of a specific quantum state by means of observations on the system.
These phenomena have their origin in the deviations from the expected exponential decay of the 
corresponding survival probability at the short-time regime \cite{CSM:PRD77}.
Various experimental confirmations on a broad list of subfields in Physics
have been reported \cite{FP:JPA08}. In particular, the occurrence of both effects has been observed in an ensemble of cold sodium atoms
trapped in an accelerated standing wave of light \cite{FGR:PRL01}.
The authors found that the decay on the number of  remaining
trapped atoms depends on the frequency of the measurements, and it could
be suppresed or enhanced in comparison with the unperturbed system.

From a theorical point of view, Kofman and Kurizki \cite{KK:N00} established that the modifications introduced to the decaying dynamics of the observed system are due to the energy spread induced by the measurements and the spectral density of the states coupled to the unstable initial state of the system.
 Facchi {\it et al.} \cite{FNP:PRL01} 
have investigated the
transition between the QZE and AZE regimes in terms of the interval separating successive measurements.
The intersection between the survival probability and the exponential decay of the state of the corresponding system according to its ``natural" rate, $\gamma_0$, defines the value
of the time
interval, $\tau^*$, for consecutive measurements which actually separate both
regimes. 

The Zeno effect has also been investigated for chemical reactions. In particular, Prezhdo \cite{P:PRL00} found evidences of quantum AZE in the photodissociation of the van der Waals (vdW) cluster HF-Na.
The process can be accelerated by an order of magnitude by controlling the loss of electronic decoherence induced by interactions between the system and the enviroment. 
The choice of the specific solvent could then be used to monitor the fragmentation process.
In a similar spirit, we have recently studied the competition between the vibrational and electronic predissociation mechanisms for the NeBr$_2$ vdW cluster \cite{SSGRM:PRL}. The branching ratio
between both fragmentation channels was found to be significantly modified by slowing down the fast vibrational predissociation pathway with successive measurements on the system.  

In this letter we address the occurrence of the Zeno effect in the context of the reaction dynamics of atom-diatom collisions studied by means of a time-dependent 
wave packet (WP) approach.
The time propagation can be written as: 

\begin{equation}
\label{timeprop}
\mid \phi(t) \rangle = e^{-iHt / \hbar} \mid \phi_0 \rangle,
\end{equation}

\noindent
where $\phi_0$ describes the value of the WP at $t=0$. The survival probability or decay
function from the initial state is simply given by:

\begin{equation}
\label{survprob}
P(t) = \mid \langle \phi_0 \mid \phi(t) \rangle\mid^2.
\end{equation}

With the measurement, the unitarity of the process, as described in 
Eq. (\ref{timeprop}),
is broken down and the quantum state describing the system ``collapses" into one of
the pointer states, $\phi_0$, associated with the measuring device. 
In particular, as a result of an ideal selective and complete measurement \cite{FP:JPA08} to investigate the decay
of the system at consecutive $\tau$ intervals, the 
state of the system at $t = N \tau$ becomes:

\begin{equation}
 \label{phiNtau}
  \mid \phi_{N \tau} \rangle = 
  \left[ \langle \phi_0 \mid \phi_{\tau} \rangle \right]^N 
  \mid \phi_0 \rangle,
\end{equation}
where the prefactor is a consequence of the 
projector $P = \mid \phi_0 \rangle \langle \phi_0 \mid$.
The corresponding survival probability can then be expressed as:

\begin{equation}
  \label{probN}
  P_{N\tau}(t) = \mid \langle \phi_0 \mid \phi_{\tau} \rangle \mid^{2N}
  \mid \langle \phi_0 \mid \phi(t) \rangle \mid^2.
\end{equation}

The comparison of the corresponding survival probabilities expressed
in Eq. (\ref{probN}) for different intervals $\tau$ and the unperturbed probability 
$P(t)$ (see Eq. (\ref{survprob})) reveals 
the possible occurrence of both QZE and AZE \cite{FNP:PRL01}.
Those $P_{N\tau}(t)$ probabilities with a slower decay in time with respect 
to $P(t)$ are usually interpreted as delayed evolutions of the 
system when we perform measurements to analyse its state, in close 
analogy to the premises stated in Zenos's paradox (``{\it stopping arrow}" approach).
The evolution of the $\phi(t)$ WP is governed by a constantly interrupted sequence
of time operators which act within the time intervals between the measurements 
performed at $t = n \tau$. 

However it is possible to conceive a  different 
rationale to explain the behaviour of the survival probabilities. 
Inspired by one of the key remarks made in Ref. \cite{FGR:PRL01}, regarding the definition of a new 
initial state after the measurement, we can devise an alternative approach in which 
the state of the system at each time $t = \tau$ interval, is employed as a reference to analyse the 
evolution of the time-propagating WP.
Thus, the survival probability after $N \tau$ measurements is written as follows: 

\begin{equation}
  \label{probN2}
  P_{N\tau}(t) = \mid \langle \phi_{(N-1) \tau} \mid \phi_{N \tau} \rangle \mid^{2}
  \mid \langle \phi_{N \tau} \mid \phi(t) \rangle \mid^2.
\end{equation}
\noindent
The prefactor in  Eq. (\ref{probN2}) can be initially understood as the constant which ensures that the 
survival probability reinitiates at the value taken at $t = (N-1) \tau$ during the 
time interval $\left[ (N-1) \tau, N \tau \right]$.
In this scheme, the WP $\phi(t)$ does not alter its time evolution (``{\it flying
arrow}" approach) which is governed by the expression given in Eq. (\ref{timeprop})
and therefore, it is possible
to express $\phi_{N \tau}$ and $\phi_{(N-1) \tau}$ in Eq. (\ref{probN2}) as a function of
$\phi_0$ according to Eq. (\ref{timeprop}).
Thus, the survival probability is expressed as:

\begin{equation}
  \label{probN3}
  P_{N\tau}(t) = \mid \langle \phi_{0} \mid \phi_{\tau} \rangle \mid^{2}
  \mid \langle \phi_{N \tau} \mid \phi(t) \rangle \mid^2.
\end{equation}

This latter approach, in which the initial WP employed as a 
reference to trace the time evolution of the system is redefined, but the evolving
WP, $\phi(t)$, as in the case of the flying arrow in Zeno's paradox, is not perturbed, 
has been highlighted by Fischer {\it et al.} \cite{FGR:PRL01}
in order to detect the QZE and AZE.
In their experiment, authors of Ref. \onlinecite{FGR:PRL01} adopted
the remaining number of atoms in the trap after the measurement as the initial condition.
The presence of the prefactor in Eq. (\ref{probN3}),
originated by the projection performed at each measurement,
thus plays an analogue role to the ansatz invoked in the experiment of Ref. \onlinecite{FGR:PRL01}
to reset the time to $t=0$.

In this letter we discuss this new approach 
for two reactions: F+HD and H+O$_2$.
The time dependent WP approach employed here has been described before \cite{GRM:JCP04} and
it is based on the description of the total Hamiltonian, $H = T + V$ in terms of mass-scaled Jacobi coordinates. 
The propagation is carried out by means of 
the symmetric split-operator method \cite{FF:JCP83} and
the application of an absorbing potential \cite{GRM:JCP04} for the outgoing fluxes at limiting regions of both reactants and products arrangements to avoid numerical reflections.  
For the corresponding $V$ of the F+HD reaction, we used the 
three dimensional SW \cite{SW:JCP96} potential energy surface (PES) while the 
DMBEIV \cite{PQBV:JPC90} PES was employed for the H+O$_2$ reactions. 
The initial WP for the F+HD case is located at the transition state, around the region of the PES at which a reactive resonance
at a collision energy of about 0.52 kcal/mol was found \cite{SSMLDL:JCP00}.
For the H+O$_2$ reaction, the initial WP started its propagation at the asymptotic region of the reactants channel.

\begin{figure*}
\includegraphics[scale=0.5]{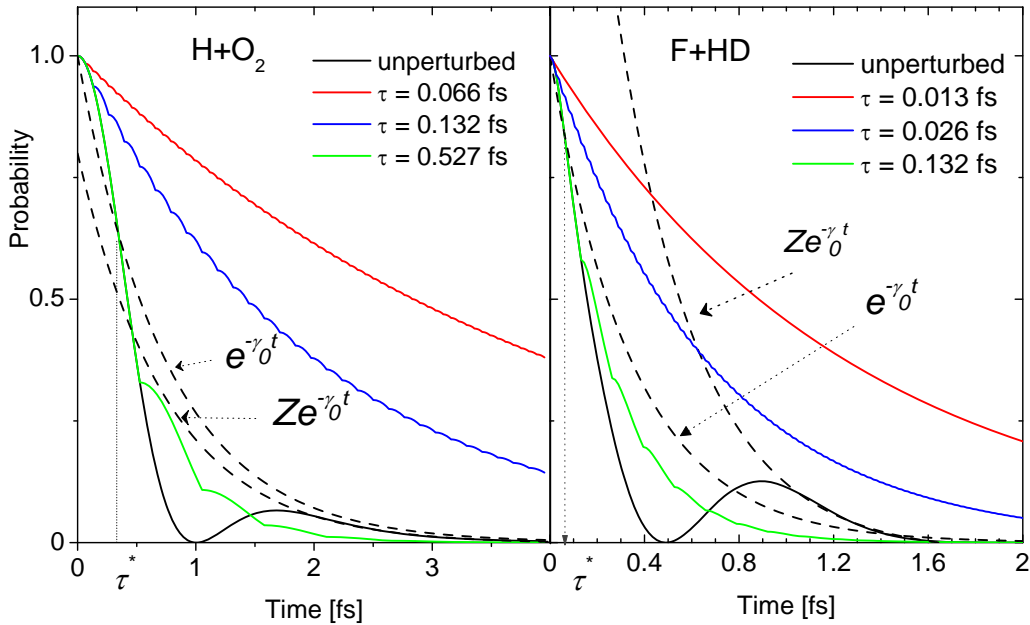}
\caption{\label{fig1} Survival probability $P(t)$ for the H+O$_2$ (left panel) and
F+HD reaction (right panel) when no redefinition of the initial WP is performed (solid black curve), and when this is done at different $t = \tau$ values. The curve with the natural decay rate $\gamma_0$ of the system observed beyond the short-time regime ${\cal Z}e^{-\gamma_0 t}$ and the decaying probability that the system would present if no deviations from the exponential behaviour were observed $e^{-\gamma_0 t}$ are included with dashed line. The crossing between the unperturbed $P(t)$ curve and $e^{-\gamma_0 t}$ at $\tau^*$ is shown with a vertical dotted line.}
\end{figure*}

The corresponding survival probability $P(t)$ as defined in Eq. (\ref{survprob}) is shown in Figure \ref{fig1} (black solid line)
for both processes. 
The curves exhibit a pronounced decay with a minimum for $t \sim 0.5$ fs, for the F+HD reaction and $t \sim 1$ fs for H+O$_2$, due to a fast dephasing between the initial WP and $\phi(t)$.
The natural exponential decay beyond the short-time regime, ${\cal Z} e^{-\gamma_0 t}$ \cite{FNP:PRL01}, is obtained from the 
tail of the secondary maximum features developed at $t = 0.9$ and 1.7 fs for the F+HD and H+O$_2$ reactions respectively. 
Both, the ${\cal Z} e^{-\gamma_0 t}$ curve and the $e^{-\gamma_0 t}$ probability which describes the exponential decay of the system from $t=0$ with this natural decay rate, are shown, with dashed lines in Figure \ref{fig1}.
In the discussion of Ref. \onlinecite{FNP:PRL01} regarding the existence of a $\tau^*$ value which separates the QZE and AZE regimes, it was found that for ${\cal Z} > 1$, the curves $e^{-\gamma_0 t}$ and $P(t)$ may not cross. Interestingly, each of the two reactions under analysis in this work, correspond to a specific case: For the F+HD, we find that ${\cal Z} > 1$, whereas for H+O$_2$, ${\cal Z} < 1$.

The $e^{-\gamma_0 t}$ curve and the unperturbed $P(t)$ probability are found
to intersect at $\tau^* \sim 0.06$ fs for the F+HD reaction and 0.32 for the H+O$_2$
reaction.  
We have calculated the survival probability according to Eq. (\ref{probN3})
for two different values of   
time intervals between consecutive measurements such that $\tau < \tau^*$
(0.013 fs and 0.026 fs for the F+HD case and 0.066 fs and 0.132 fs for the H+O$_2$). 
These $P_{\tau}(t)$ decay probabilies, shown in Fig. \ref{fig1}, extend noticeably longer 
in time than the original $P(t)$: 
they can be then interpreted as a manifestation of the QZE.

Alternatively, the same procedure is applied to measurements spaced in time
a value $\tau > \tau^*$. In particular, we have considered $\tau =$ 0.132 fs for
F+HD and 0.527 fs for H+O$_2$.
The crossing between the corresponding $P_{\tau}(t)$ survival probabilities 
and the unperturbed  $P(t)$ curves for each reaction finally manifests 
much faster decays to zero when we operate on the system, 
an evidence of the AZE.    

It is worth mentioning that the survival probabilities obtained after the 
operational procedure described in Eqs. (\ref{probN2}) and (\ref{probN3}) exhibit a profile 
in which the recurrence observed for the unperturbed $P(t)$ probabilities
is completely lost.
Those $P_{N \tau}(t)$ probabilities, on the other hand, exhibit an oscillating structure
with values at each $t = n \tau$ time interval which follow perfectly, as discussed in Ref. \onlinecite{FNP:PRL01},
the exponentially decay governed by the $\gamma(\tau)$ rate. 


Whereas the possible differences observed on the survival probabilities
obtained as in  Eq. (\ref{probN}) with respect to the unperturbed probability
$P(t)$ could certainly be explained in terms of variations of the evolution of the
system from its initial state (at least in the subspace defined by the measurement
\cite{FP:JPA08}), the interpretation for the results obtained via the new approach proposed
in this letter
involves more subtle considerations.
Under the assumption of the projected state
after the measurement  as the new starting point for the propagation,
the actual 
evolution of the system, here represented by the time dependent WP, is not affected at all: the WP remains unaltered and no differences in the observed dynamics is thus expected after the measurements on the system.
In particular, the observed QZE could be understood, not as a hindered evolution, but as a consequence of the update of the reference used to measure the overall change of 
the state of the system.

On the other hand, the occurrence of the AZE seems to be subject to
some geometrical considerations, as stated by Facchi {\it et al.} \cite{FNP:PRL01}: the definition of $\tau^*$, the limiting value between the QZE and AZE of the time intervals for the measurement, is based on the expected exponential decay of the survival probability. 
In this sense, the probabilities shown in Fig. \ref{fig1} exhibit such an exponential behaviour once a secondary maximum is developed, in a similar way as the examples discussed 
in Ref. \onlinecite{FNP:PRL01}.
The $\tau^*$ value is then required to be sufficiently large to ensure that the $P(t)$ function 
is low enough at the moment of the measurement; thus the ulterior evolution will yield probabilities which do not exceed the limit imposed by the exponentially decaying behaviour followed beyond the short-time regime. 

Our findings are consistent with a short-time dynamics. First of all, the time scale
of Fig. \ref{fig1} is in line with this specific regime, with most of the interesting
features occurring well below 2 fs for both the H+O$_2$ and F+HD reactions here
investigated. 
Furthermore, one might wonder if the survival probabilities
as expressed in Eqs. (\ref{probN}) and (\ref{probN3}) are, in essence,
formally equivalent, despite of the fact that the corresponding $\phi(t)$ WPs follow
different time evolutions. 
This seems to be the case if we could apply the definition of the 
$\phi_{N \tau}$ function given in Eq. (\ref{phiNtau}) to expand the 
$P_{N \tau}(t)$ probability of the ``{\it flying arrow}" approach 
(see Eq. (\ref{probN3})). 
The redefined initial state at each $t = \tau$ would be then a rescaling of the state of the system at $t = 0$, $\phi_0$, with a prefactor which reduces the
corresponding amplitude.
The above mentioned short-time regimes seem to be the 
restriction for the strict validity of such approaches.  

Since our analysis is enterely based on the projection of the initial state $\phi_0$ and their corresponding time-updated expressions $\phi_{\tau}$, one might wonder about the possible universality of the occurrence of the QZE and AZE scenarios here described. 
In our calculations
an ample set of different locations and average incident momenta, $p_0$, for the initial WP were found to lead to similar results, but, for the F+HD reaction, 
the most favourable cases were those in which the WP was located near to the transition state, the choice discussed in this paper.

The dependence of the occurrence of both QZE and AZE on the parameters of the initial 
WP can be better analyzed in the one-dimensional case. 
For an initial state described by a Gaussian WP in free space, $V(x) = 0$, the survival 
probability in the short time limit can be expressed as \cite{SSGRM:sub}:

\begin{equation}
\label{survP1D}
P(t) = \left( 1 - {t^2 \over 8 t_c^2} \right)
e^{-2(p_0/p_s)^2(t^2 / 8 t_c^2)},
\end{equation}
\noindent
where $p_s = \hbar / 2 \sigma_0$ is the spreading momentum and $t_c = 2m \sigma_0^2 / \hbar^2$ is a characteristic time with $\sigma_0$ being the
initial spreading of the WP.
If the translational and spreading motion of the WP are such that $p_0 / p_s \ll 2 t_c / t$, then
Eq. (\ref{survP1D}) is written as:

\begin{equation}
\label{survP1Db}
P(t) = 1 - \left[ 1 + 2 \left( p_0 \over p_s \right)^2 \right] {t^2 \over 8 t_c^2}, 
\end{equation}
\noindent
which displays the characteristic quadratic behaviour with time of Zeno's scenario 
\cite{FP:JPA08}.
For any other relation between $p_0$ and $p_s$, the survival probability decays too fast
to observe the QZE.
In the above Eq. (\ref{survP1Db}), if the translation momentum of the WP is too
small to be neglected in comparison with the spreading motion, $p_0 \approx 0$,
we find that the $\tau^*$ time which separates the QZE and AZE is given by
$\tau^* = 2 \sqrt{2} \, t_c$.


According to this, the QZE will be observed whenever the measurements are performed 
before the WP has entered the time-scales of linear spreading.
If, on the contrary, the WP has some non-zero initial momentum, we find that:

\begin{equation}
\tau^* = {2 \sqrt{2} \, t_c \over \sqrt{1+2(p_0/p_s)^2}}.
\end{equation}

Our present study revisits Zeno's paradox for time-propagating WPs in the context of atom-diatom 
reactions. 
In order to understand the effect of frequent measurements on the survival probabilities, we
propose a novel approach: the redefinition of the initial state used as a reference to calculate the decay of the state of the system.
Thus, apparently counterintuitive explanations regarding the evolution of the
time evolving WPs are not required. A possible extention to the present study could include resonances and quasibound states, 
examples of time-decaying features within the context of molecular processes treated here.






This work has been supported by the Spanish Ministerio de Ciencia e INnovaci\'on (MCIN) under projects FIS2010-18132 and FIS2010-22082, and grant CSD2009-00038 within the Program CONSOLIDER-INGENIO, and by Comunidad Aut\'onoma de Madrid under Grant S-2009/MAT/1467. ASS would also like to thank the MCIN for a ``Ram\'on y Cajal'' Research Fellowship.



\end{document}